# In-plane chemical pressure essential for superconductivity in $BiCh_2$-based (Ch: S, Se) layered structure

Yoshikazu Mizuguchi[1*], Akira Miura[2], Joe Kajitani[1], Takafumi Hiroi[1], Osuke Miura[1], Kiyoharu Tadanaga[2], Nobuhiro Kumada[3], Eisuke Magome[4], Chikako Moriyoshi[4], Yoshihiro Kuroiwa[4]

1. Department of Electrical and Electronic Engineering, Tokyo Metropolitan University, 1-1, Minami-osawa, Hachioji 192-0397, Japan.
2. Faculty of Engineering, Hokkaido University, Kita-13, Nishi-8, Kita-ku, Sapporo 060-8628 Japan.
3. Center for Crystal Science and Technology, University of Yamanashi, 7-32 Miyamae, Kofu 400-8511 Japan.
4. Department of Physical Science, Hiroshima University, 1-3-1 Kagamiyama, Higashihiroshima, Hiroshima 739-8526 Japan.

**Abstract**

$BiCh_2$-based superconductors (Ch: S, Se) are a new series of layered superconductor. However, mechanisms for the emergence of superconductivity in $BiCh_2$-based superconductors have not been clarified. In this study, we have investigated crystal structure of two series of optimally-doped $BiCh_2$-based superconductors, $Ce_{1-x}Nd_xO_{0.5}F_{0.5}BiS_2$ and $LaO_{0.5}F_{0.5}Bi(S_{1-y}Se_y)_2$, using powder synchrotron x-ray diffraction in order to reveal the relationship between crystal structure and superconducting properties of the $BiCh_2$-based family. We have found that an enhancement of in-plane chemical pressure would commonly induce bulk superconductivity in both systems. Furthermore, we have revealed that superconducting transition temperature for $REO_{0.5}F_{0.5}BiCh_2$ superconductors could universally be determined by degree of in-plane chemical pressure.

Most of superconductors with a high transition temperature ($T_c$) possess a layered crystal structure. Typical examples of layered high-$T_c$ superconductor are Cu-oxide [1] and Fe-based superconductors [2]. They have a crystal structure composed of an alternate stacking of superconducting layers ($CuO_2$ or FeAs layers) and spacer layers. The spacer layers are electrically insulating and essential for the emergence of low-dimensional electronic states within the superconducting layers, which is sometimes responsible for the emergence of unconventional paring mechanisms. One of the attractive features of those layered superconductors is a wide variation of crystal structure. Various kinds of new superconductors can be designed by stacking common superconducting layers and various types of spacer layers, and their superconducting properties depend on types of the spacer layers. In fact, many Fe-based superconductors with various types of spacer layer have been discovered, and high $T_c$ was achieved in several Fe-based superconductors [2-8].

With a wide variation of related layered superconductors with various structures and properties, one can extract a crystal structure parameter essential for the emergence of superconductivity in the layered family. In the Fe-based family, it was found that $T_c$ can be estimated by a crystal structure parameter such as an As-Fe-As bond angle [9] or an anion height from Fe square lattice (anion = As, P, Se, and Te) [10]. Changes in those parameters strongly affect configurations of Fermi surfaces and paring symmetry, and superconducting properties of Fe-based superconductors [11]. In fact, clarification of a crystal structure correlating with superconducting properties is one of the most important challenges for understanding the mechanisms of superconductivity in a new series of layered superconductor. In addition, it gives a direct strategy for designing new layered superconductors with a higher $T_c$.

In 2012, we reported the discovery of novel layered superconductors with a crystal structure composed of an alternate stacking of common $BiS_2$ superconducting layers and various spacer layers [12,13]. Band calculations suggested that the parent phase of $BiS_2$-based superconductors is an insulator with $Bi^{3+}$. Superconductivity is induced when electron carriers are doped into the $BiS_2$ layers (within Bi-6p orbitals) by element substitution at the spacer layers [12,14]. For example, a parent compound $REOBiS_2$ becomes a superconductor by a partial substitution of O at the spacer layers by F (namely, $REO_{1-x}F_xBiS_2$) [13]. So far, 12 kinds of parent compounds of $BiS_2$-based superconductor have been discovered: $REOBiS_2$ (RE = La [13], Ce [15], Pr [16], Nd [17], Sm [18], Yb [19], and Bi [20,21]), $AFBiS_2$ (A = Sr [22,23], Eu [24]), $Bi_6O_8S_5$ [12], $Bi_3O_2S_3$ [25], and $Eu_3F_4Bi_2S_4$ [26]. In addition, superconductivity was observed in a $BiSe_2$-based compound $LaO_{0.5}F_{0.5}BiSe_2$ as well [27]. Therefore, exploration for new $BiCh_2$-based (Ch:

S, Se) layered superconductors and studies with an aim to elucidate the mechanisms of superconductivity in the $BiCh_2$ layered compounds have attracted researchers in the field of condensed matter physics.

As mentioned above, electron-carrier doping is needed for the emergence of superconductivity in the $BiCh_2$ family. On the other front, some reports have suggested that optimization of crystal structure is also important for the emergence of superconductivity in the $BiCh_2$ family. One of those is a high pressure (HP) effect on $T_c$. For example, $LaO_{0.5}F_{0.5}BiS_2$ shows dramatic changes in superconductivity under HP. $LaO_{0.5}F_{0.5}BiS_2$ does not show bulk superconductivity while it shows a filamentary (weak) superconductivity signal with a $T_c$ of 2.5 K [13]. With applying HP, $LaO_{0.5}F_{0.5}BiS_2$ becomes a bulk superconductor and $T_c$ largely increases from 2.5 K to over 10 K [13,28-32]. In addition, an enhancement of $T_c$ under HP was observed in other $REO_{0.5}F_{0.5}BiS_2$ superconductors as well [32]. These facts strongly suggest that superconducting properties of $BiCh_2$-based superconductors are correlated with changes in crystal structure as revealed in the Fe-based family [9,10].

Furthermore, our recent studies on isovalent-substitution effects to superconductivity have suggested that optimization of crystal structure is important for the emergence of bulk superconductivity and a higher $T_c$ in optimally-doped $REO_{0.5}F_{0.5}BiCh_2$. One of the examples is $Ce_{1-x}Nd_xO_{0.5}F_{0.5}BiS_2$ [33]. Since valence of Ce and Nd are basically 3+ in this crystal structure, electron carriers in these compounds are essentially the same: a formal valence of Bi is 2.5+. However, bulk superconductivity is induced by a systematic substitution of Ce by Nd, and $T_c$ increased with increasing Nd concentration ($x$) as shown in Fig. 1a. The emergence of superconductivity was explained by uniaxial lattice shrinkage along the $a$-axis and optimization of a lattice shrinkage ratio $c$ / $a$ [33]. The other example of the isovalent-substitution systems is $LaO_{0.5}F_{0.5}Bi(S_{1-y}Se_y)_2$ [34]. In $LaO_{0.5}F_{0.5}Bi(S_{1-y}Se_y)_2$, the $S^{2-}$ site within the superconducting layers is systematically substituted by $Se^{2-}$. Therefore, the formal valence of $Bi^{2.5+}$ should not be changed by Se substitution. Bulk superconductivity is induced by Se substitution, and $T_c$ increases with increasing Se concentration ($y$) as shown in Fig. 1b. In this system, Se substitution enhances metallic conductivity, and bulk superconductivity is induced by an expansion of lattice volume by Se substitution [34].

On the basis of these experimental facts in HP studies on $REO_{0.5}F_{0.5}BiS_2$ [13,28-32] and isovalent-substitution studies on $Ce_{1-x}Nd_xO_{0.5}F_{0.5}BiS_2$ and $LaO_{0.5}F_{0.5}Bi(S_{1-y}Se_y)_2$ [33,34], there is no doubt that superconducting properties of the $BiCh_2$ family correlate with not only electron carrier concentration but also their crystal structure. However, universal relationship between crystal structure and the emergence of superconductivity (and their $T_c$) in the $BiCh_2$ family has not been clarified yet. In this study, we aim to clarify a crystal structure which directly correlates with the emergence of superconductivity and $T_c$ of the $BiCh_2$ family. Fortunately, we have two isovalent-substitution systems of $Ce_{1-x}Nd_xO_{0.5}F_{0.5}BiS_2$ and $LaO_{0.5}F_{0.5}Bi(S_{1-y}Se_y)_2$, which exhibit a similar superconductivity phase diagram to each other as shown in Fig. 1. In order to discuss the changes in crystal structure parameters and their relationship to superconductivity, we have performed powder synchrotron x-ray diffraction (XRD) and Rietveld refinement for $Ce_{1-x}Nd_xO_{0.5}F_{0.5}BiS_2$ and $LaO_{0.5}F_{0.5}Bi(S_{1-y}Se_y)_2$.

Here, we show that an enhancement of in-plane chemical pressure within the superconducting Bi-Ch plane would be a crystal structure parameter, which commonly explain the emergence of superconductivity in $REO_{0.5}F_{0.5}BiCh_2$ compounds. Furthermore, we show that $T_c$ for $REO_{0.5}F_{0.5}BiCh_2$ superconductors could universally be determined by degree of in-plane chemical pressure.

## Results

### Evolution of crystal structure parameters.

We have performed powder synchrotron XRD for $Ce_{1-x}Nd_xO_{0.5}F_{0.5}BiS_2$ and $LaO_{0.5}F_{0.5}Bi(S_{1-y}Se_y)_2$ and performed Rietveld refinement. Typical Rietveld refinement profiles for $x = 0.6$ and $y = 0.5$ are displayed in supplementary information (Figure S1). Although tiny impurity phases of RE fluorides were detected, all the XRD peaks of the main phase were refined using a tetragonal *P*4/*nmm* space group. Obtained crystal structure parameters are plotted as a function of $x$ (or $y$) in Fig. 2: the crystal structure data are listed in supplementary information (Table S2). In Fig. 2, the data points for the samples showing bulk superconductivity (Bulk SC) are highlighted with orange-filled circles. If a crystal structure parameter for $Ce_{1-x}Nd_xO_{0.5}F_{0.5}BiS_2$ and $LaO_{0.5}F_{0.5}Bi(S_{1-y}Se_y)_2$ changed similarly to each other, the crystal structure parameter should be essential for the emergence of superconductivity in these series because these two series exhibit a similar superconductivity phase diagram as a function of $x$ or $y$ as introduced in Fig. 1.

Figure 2a and 2b show the $x$ (or $y$) dependences of lattice constants of $a$ and $c$ for $Ce_{1-x}Nd_xO_{0.5}F_{0.5}BiS_2$ and $LaO_{0.5}F_{0.5}Bi(S_{1-y}Se_y)_2$, respectively. In $Ce_{1-x}Nd_xO_{0.5}F_{0.5}BiS_2$, both lattice constants decrease with increasing $x$ due to an increase in concentration of $Nd^{3+}$ (112 pm, assuming coordination number of 8), which is smaller than $Ce^{3+}$ (114 pm). In contrast, both lattice constants increase with increasing $y$ in $LaO_{0.5}F_{0.5}Bi(S_{1-y}Se_y)_2$ due to an increase in concentration of $Se^{2-}$ (198 pm, assuming coordination number of 6), which is larger than $S^{2-}$ (184 pm). These contrasting changes in lattice constants suggest that the evolution of superconductivity in these two series cannot be explained by a simple lattice contraction or expansion.

Figure 2c shows the *x* (or *y*) dependences of Ch1-Bi-Ch1 angle for $Ce_{1-x}Nd_xO_{0.5}F_{0.5}BiS_2$ and $LaO_{0.5}F_{0.5}Bi(S_{1-y}Se_y)_2$. The Ch1-Bi-Ch1 angle is an indicator of flatness of Bi-Ch1 plane. Since electrons within the Bi-6p orbitals hybridized with the Ch-p orbitals (S-3p or Se-4p) are essential for the emergence of superconductivity in $BiCh_2$-based superconductors, a flatter Bi-Ch1 plane simply seems to be better for the emergence of bulk superconductivity with a higher $T_c$. In fact, a previous study on crystal structure of $CeO_{1-x}F_xBiS_2$ single crystals with different F concentrations indicated that a flatter Bi-S1 plane resulted in higher superconducting properties [35]. In $Ce_{1-x}Nd_xO_{0.5}F_{0.5}BiS_2$, the Ch1-Bi-Ch1 angle approaches 180 degree with increasing *x*, indicating that the Bi-S1 plane becomes flatter, and superconductivity is induced with Nd substitution. However, the Se concentration dependence of Ch1-Bi-Ch1 angle for $LaO_{0.5}F_{0.5}Bi(S_{1-y}Se_y)_2$ exhibits a contrasting behavior to that observed with Nd substitution effect in $Ce_{1-x}Nd_xO_{0.5}F_{0.5}BiS_2$. With increasing *y*, the Ch1-Bi-Ch1 angle decreases: the Bi-Ch1 plane becomes distorted. Therefore, the Ch1-Bi-Ch1 angle (flatness of the Bi-Ch1 plane) cannot commonly explain the evolution of superconductivity in $BiCh_2$-based superconductors. The contrasting changes in the in-plane structure (flatness) may be due to the difference in $Ch^{2-}$ ions at the Ch1 site. Figure 2d shows the *y* dependence of occupancy of Se at the Ch1 site. As it is clear in Fig. 2d, Se ions selectively occupy the in-plane Ch1 site. At $x = 0.5$, approximately 90% of Ch1 site is occupied with Se. Recently, similar site selectivity of Se in $LaO_{1-x}F_xBiSSe$ single crystals were reported [36]. The preferential occupation of $Se^{2-}$ with a larger ionic radius within the two-dimensional Bi-Ch1 plane may cause distortion.

Figure 2e shows the *x* (or *y*) dependences of three kinds of Bi-Ch distance. As depicted in a right-side image of the $BiCh_2$ layer, a Bi ion is coordinated by six Ch ions. The shortest Bi-Ch distance is the Bi-Ch2 distance toward a *c*-axis direction. The Bi-Ch2 distance for $Ce_{1-x}Nd_xO_{0.5}F_{0.5}BiS_2$ slightly decreases with increasing *x*. In contrast, the Bi-Ch2 distance for $LaO_{0.5}F_{0.5}Bi(S_{1-y}Se_y)_2$ increases with increasing *y*. Therefore, the Bi-Ch2 distance should not be essential for the evolution of superconductivity within these two series. Next, we discuss the in-plane Bi-Ch1 distance. The Bi-Ch1 (in-plane) distance for $Ce_{1-x}Nd_xO_{0.5}F_{0.5}BiS_2$ exhibits a decrease with increasing *x*. In contrast, the Bi-Ch1 (in-plane) distance for $LaO_{0.5}F_{0.5}Bi(S_{1-y}Se_y)_2$ increases with increasing *y*. These facts suggest that the in-plane Bi-Ch1 distance itself cannot explain the evolution of superconductivity. The longest Bi-Ch distance is Bi-Ch1 (inter-plane), which is roughly corresponding to the inter-layer distance of two $BiCh_2$ layers. The Bi-Ch1 (inter-plane) distance for $Ce_{1-x}Nd_xO_{0.5}F_{0.5}BiS_2$ does not exhibit a remarkable change upon Nd substitution while that for $LaO_{0.5}F_{0.5}Bi(S_{1-y}Se_y)_2$ clearly increases with increasing $y$. Therefore, the Bi-Ch1 (inter-plane) distance cannot commonly explain the evolution of superconductivity.

Although we expected that a Ch1-Bi-C1 angle or some Bi-Ch distance exhibit a common behavior in $Ce_{1-x}Nd_xO_{0.5}F_{0.5}BiS_2$ and $LaO_{0.5}F_{0.5}Bi(S_{1-y}Se_y)_2$, we could not extract any clear correlation between them. However, we assumed that the in-plane Bi-Ch1 distance should correlate with the evolution of superconductivity to a certain extent because the superconductivity is induced within the Bi-Ch1 plane, and their superconducting properties dramatically change by changing the crystal structure without changes in electron carrier concentration (F concentration). Therefore, we introduce a concept of *in-plane chemical pressure* to discuss the relationship between in-plane structure and evolution of superconductivity in $Ce_{1-x}Nd_xO_{0.5}F_{0.5}BiS_2$, $LaO_{0.5}F_{0.5}Bi(S_{1-y}Se_y)_2$, and other $REO_{0.5}F_{0.5}BiCh_2$ superconductors.

**Emergence of superconductivity by in-plane chemical pressure.**

Figure 3a shows schematic images of compression or expansion of Bi-Ch plane by Nd or Se substitution. In the case of $Ce_{1-x}Nd_xO_{0.5}F_{0.5}BiS_2$, Bi-Ch1 planes are compressed according to a decrease in volume of spacer layers with increasing Nd concentration ($x$). The compression of Bi-Ch1 plane results in an enhancement of packing density of $Bi^{2.5+}$ and $S^{2-}$ ions within the superconducting plane: this is so-called *in-plane chemical pressure*. In the case of $LaO_{0.5}F_{0.5}Bi(S_{1-y}Se_y)_2$, the in-plane Bi-Ch1 distance increases with increasing occupancy of Se at the Ch1 site. However, the expansion of the in-plane Bi-Ch1 distance is smaller than that simply expected from the difference in ionic radius of $S^{2-}$ and $Se^{2-}$ because the component of the spacer layer (LaO) never changes in $LaO_{0.5}F_{0.5}Bi(S_{1-y}Se_y)_2$. Therefore, the packing density of $Bi^{2.5+}$ and $Ch^{2-}$ ions within the superconducting plane is enhanced. This situation is actually similar to the enhancement of in-plane chemical pressure in $Ce_{1-x}Nd_xO_{0.5}F_{0.5}BiS_2$. In order to compare degree of in-plane chemical pressure of two series, we define a value of in-plane chemical pressure using equation (1).

$$\text{In-plane chemical pressure} = (R_{\text{Bi}} + R_{\text{Ch1}}) / \text{Bi-Ch1(in-plane)} \quad (1)$$

$R_{\text{Bi}}$ is an ionic radius of $Bi^{2.5+}$. Here, we assumed that the ionic radius of $Bi^{2.5+}$ is 104.19

pm, which is obtained from the average of the six Bi-S bonds (four in-plane Bi-S1 bonds, one inter-plane Bi-S1 bond, and one Bi-S2 bond) determined from a crystal structure analysis with a single crystal of $LaO_{0.54}F_{0.46}BiS_2$ [37]. $R_{Ch1}$ is an ionic radius of chalcogen at the Ch1 site: 184 and 198 pm for $S^{2-}$ and $Se^{2-}$, respectively. In the case of $LaO_{0.5}F_{0.5}Bi(S_{1-y}Se_y)_2$, we calculated an average value for $R_{Ch1}$ using occupancy of Se at the Ch1 site. The Bi-Ch1 (in-plane) is the data obtained from Rietveld refinement (Fig. 2e). We note that chemical pressure derived from ionic radii is a simple estimation, and it cannot describe exact orbital overlap between Bi and Ch. However, the assumption looks very useful to discuss the relationship between crystal structure and superconductive properties of these systems.

The calculated in-plane chemical pressure is plotted as a function of $x$ (or $y$) in Fig. 3b. For both systems, in-plane chemical pressure increases with increasing $x$ (or $y$). Surprisingly, the chemical pressure, at which bulk superconductivity is induced, is almost the same (above ~1.011) in between $Ce_{1-x}Nd_xO_{0.5}F_{0.5}BiS_2$ and $LaO_{0.5}F_{0.5}Bi(S_{1-y}Se_y)_2$. On the basis of these experimental facts, we suggest that the emergence of bulk superconductivity in both $Ce_{1-x}Nd_xO_{0.5}F_{0.5}BiS_2$ and $LaO_{0.5}F_{0.5}Bi(S_{1-y}Se_y)_2$ systems can commonly be explained by the increase of in-plane chemical pressure. The enhancement of in-plane chemical pressure should enhance the overlaps of Bi-6p and Ch-p orbitals. The enhancement of orbital overlaps would result in an enhancement of metallic conductivity and an inducement of bulk superconductivity in $REO_{0.5}F_{0.5}BiCh_2$ family. In the discussion part, we discuss the relationship between $T_c$ and in-plane chemical pressure within the Bi-Ch1 plane.

## Discussion

To discuss the relationship between $T_c$ and in-plane chemical pressure, we plotted $T_c$ for $Ce_{1-x}Nd_xO_{0.5}F_{0.5}BiS_2$ and $LaO_{0.5}F_{0.5}Bi(S_{1-y}Se_y)_2$ as a function of in-plane chemical pressure in Fig. 4. To obtain a general tendency in $REO_{0.5}F_{0.5}BiCh_2$ superconductors, we added data points for $Nd_{0.8}Sm_{0.2}O_{0.5}F_{0.5}BiS_2$ and $Nd_{0.6}Sm_{0.4}O_{0.5}F_{0.5}BiS_2$ [analyzed in this study], $NdO_{0.5}F_{0.5}BiS_2$ single crystal [38], $PrO_{0.5}F_{0.5}BiS_2$ [16], $SmO_{0.5}F_{0.5}BiS_2$ single crystal [18], and $LaO_{0.5}F_{0.5}BiSe_2$ ($y$ = 1) single crystal [39]. Interestingly, data points for all the $REO_{0.5}F_{0.5}BiS_2$-type are located on a single slope highlighted by a blue region. Notably, the data points for $CeO_{0.5}F_{0.5}BiS_2$ and $SmO_{0.5}F_{0.5}BiS_2$, which do not exhibit superconductivity, locate on the left region of the boundary suggested in Fig. 4 (in-plane chemical pressure < 1.011). These facts suggest that the emergence of superconductivity and $T_c$ of $REO_{0.5}F_{0.5}BiS_2$-type simply depend on degree of in-plane chemical pressure. If we consider the BCS (Bardeen-Cooper-Schrieffer) theory with electron-phonon mechanisms [40], the enhancement of $T_c$ would be explained by an increase of phonon frequency and/or an enhancement of density of state at the Fermi level. Generally in metals, an increase in overlap of orbitals should decrease density of state due to an increase of band width. Therefore, an increase in density of state could not simply explain the enhancement of $T_c$ in this series. Although phonon frequency might be enhanced with increasing in-plane chemical pressure, further experimental and theoretical investigations are needed to elucidate the mechanisms of the enhancement of $T_c$ with increasing in-plane chemical pressure. Notably, $T_c$ for $LaO_{0.5}F_{0.5}Bi(S_{1-y}Se_y)_2$ and $LaOBiSe_2$, indicated with a red region, are clearly located on a lower side of that for the $REO_{0.5}F_{0.5}BiS_2$-type series. As an important fact, evolutions of $T_c$ in $REO_{0.5}F_{0.5}BiS_2$ and $LaO_{0.5}F_{0.5}Bi(S_{1-y}Se_y)_2$ do not correspond and exhibit different curves. For the $LaO_{0.5}F_{0.5}Bi(S_{1-y}Se_y)_2$ compounds, in-plane chemical pressure is relatively stronger than that in $REO_{0.5}F_{0.5}BiS_2$, but their $T_c$ is clearly lower than those of $REO_{0.5}F_{0.5}BiS_2$. As described above, Se preferably occupies the in-plane Ch1 site. Therefore, it can be considered that superconductivity in $LaO_{0.5}F_{0.5}Bi(S_{1-y}Se_y)_2$ is induced in a Bi-(S,Se) or a Bi-Se plane. If we assume the BCS theory, the lower $T_c$ in Bi-Se plane can be explained by a lower phonon frequency in Bi-Se plane than that in Bi-S plane due to the larger atomic number of Se. In order to confirm the assumption above and mechanisms of superconductivity in $BiCh_2$-based family, we need to obtain other $REO_{0.5}F_{0.5}BiSe_2$ superconductors like $CeO_{0.5}F_{0.5}BiSe_2$ or $NdO_{0.5}F_{0.5}BiSe_2$ and add data points on an extension of the $LaO_{0.5}F_{0.5}Bi(S_{1-y}Se_y)_2$ series. As a conclusion with Fig. 4, we expect that a higher $T_c$ would be obtained with Bi-S plane with a higher in-plane chemical pressure in $REO_{0.5}F_{0.5}BiCh_2$-type compounds.

At the end, we briefly discuss about the HP phase of $LaO_{0.5}F_{0.5}BiS_2$ which shows the highest $T_c$ among $BiCh_2$-based superconductors [13,28-32]. T. Tomita *et al.* reported that the crystal structure of $LaO_{0.5}F_{0.5}BiS_2$ above 0.7 GPa was monoclinic [31]. In the monoclinic structure, the Bi-S1 plane is distorted into a zigzag chain with a shorter Bi-S1 distance of 2.72 Å; zigzag chains are connected with a longer Bi-S1 distance of 3.03 Å to each other. In fact, we reported in-plane anisotropy of upper critical field within the Bi-S1 plane in a HP phase of $LaO_{0.5}F_{0.5}BiS_2$, implying the emergence of quasi-one-dimensional superconducting states within the distorted Bi-S1 plane of a HP phase of $LaO_{0.5}F_{0.5}BiS_2$ [41]. If we simply calculate the value of chemical pressure using a Bi-S1 distance of 2.72 Å with the equation (1), we obtain chemical pressure of 1.06. (In this case, obtained chemical pressure is not in-plane chemical pressure but quasi-one-dimensional chemical pressure.) Chemical pressure of 1.06 and $T_c$ of 10 K give a data point locating on a rough extension of the $T_c$-chemical pressure curve for the other $REO_{0.5}F_{0.5}BiS_2$ superconductors discussed in Fig. 4. This might indicate that the enhancement of quasi-one-dimensional chemical pressure, not in-plane chemical pressure, would be essential for the evolution of superconductivity in $REO_{0.5}F_{0.5}BiCh_2$ compounds. To further discuss and understand the relationship between chemical pressure and superconductivity in the $BiCh_2$ family, investigations of crystal structure and superconducting properties for new $BiS_2$-, $BiSe_2$-based and/or $BiTe_2$-based compounds with various spacer layers are needed.

In summary, we have analyzed crystal structure of optimally-doped $BiCh_2$-based superconductors $Ce_{1-x}Nd_xO_{0.5}F_{0.5}BiS_2$ and $LaO_{0.5}F_{0.5}Bi(S_{1-y}Se_y)_2$ using powder synchrotron XRD to reveal relationship between crystal structure and superconducting properties in $BiCh_2$-based layered compounds. We have found that the enhancement of in-plane chemical pressure could commonly induce bulk superconductivity in both systems. Furthermore, we revealed that $T_c$ for the $REO_{0.5}F_{0.5}BiCh_2$ superconductors could be determined by degree of in-plane chemical pressure and a kind of Ch element composing superconducting Bi-Ch planes. In addition, on the basis of discussion on the relationship between chemical pressure and superconductivity in $REO_{0.5}F_{0.5}BiCh_2$ compounds and a monoclinic phase (high pressure phase) of $LaO_{0.5}F_{0.5}BiS_2$, we suggest a possibility that the enhancement of quasi-one-dimensional chemical pressure within a Bi-Ch chain, not in-plane (two-dimensional) chemical pressure, would be essential for a higher $T_c$ in $BiCh_2$-based superconductors.

## Methods

Polycrystalline samples of $Ce_{1-x}Nd_xO_{0.5}F_{0.5}BiS_2$ and $LaO_{0.5}F_{0.5}Bi(S_{1-y}Se_y)_2$ used in this study were prepared using solid state reaction. The detailed synthesis procedures were described in previous reports [33,34]. Powder synchrotron X-ray powder diffraction measurements were performed at room temperature at the BL02B2 experimental station of SPring-8 (JASRI; Proposal No. 2014B1003 and 2014B1071). The wavelength of the radiation beam was 0.49542(4) Å. We have performed the Rietveld refinement (RIETAN-FP [42]) for $Ce_{1-x}Nd_xO_{0.5}F_{0.5}BiS_2$ and $LaO_{0.5}F_{0.5}Bi(S_{1-y}Se_y)_2$ using a typical structure model of $REOBiCh_2$-based superconductor with a tetragonal space group of $P4/nmm$ [36,37]. Contributions from impurity phases of rare-earth fluorides ($REF_3$) and/or $Bi_2S_3$ are included in Rietveld refinement. In addition, we have analyzed a crystal structure for two related compounds of $Nd_{0.8}Sm_{0.2}O_{0.5}F_{0.5}BiS_2$ and $Nd_{0.6}Sm_{0.4}O_{0.5}F_{0.5}BiS_2$ to enrich the discussion part [33]. The obtained crystal structure parameters are summarized in Table S1. The schematic images of crystal structure were drawn using VESTA [43].

**Acknowledgements**

The authors would like to thank Dr. N. L. Saini of Sapienza University of Rome, Dr. K. Kuroki of Osaka University and Dr. Y. Takano of National Institute for Materials Science for fruitful discussion. This work was partly supported by Grant-in-Aid for Young Scientist (A): 25707031 and Grant-in-Aid for challenging Exploratory Research: 26600077. The synchrotron x-ray diffraction were performed under proposals of JASRI; Proposal No. 2014B1003 and 2014B1071.

**Author contributions**

Y.M. and A.M. planned the research. Y.M., J.K., T.H and O.M. prepared polycrystalline samples and studied superconducting properties. Y.M., A.M., K.T., N.K., E.M., C.M. and Y.K. performed synchrotron XRD and crystal structure analysis. Y.M. wrote the manuscript.

## References

[1] Bednorz, J. G. & Müller, K. A. Possible high $T_c$ superconductivity in the Ba–La–Cu–O system. *Z. Physik B Condensed Matter* **64**, 189-193 (1986).

[2] Kamihara, Y. *et al.*, Iron-Based Layered Superconductor La[$O_{1-x}F_x$]FeAs (x=0.05–0.12) with $T_c$ = 26 K. *J. Am. Chem. Soc.* **130**, 3296–3297 (2008).

[3] Chen, X. H. *et al.*, Superconductivity at 43 K in $SmFeAsO_{1-x}F_x$. *Nature* **453**, 761-762 (2008).

[4] Ren, Z. A. *et al.*, Superconductivity at 55 K in Iron-Based F-Doped Layered Quaternary Compound Sm[$O_{1-x}F_x$] FeAs. *Chinese Phys. Lett.* **25**, 2215 (2008).

[5] Rotter, M., Tegel, M. & Johrendt, D. Superconductivity at 38 K in the Iron Arsenide $(Ba_{1-x}K_x)Fe_2As_2$. *Phys. Rev. Lett.* **101**, 107006(1-4) (2008).

[6] Wang, X.C. *et al.*, The superconductivity at 18 K in LiFeAs system. *Solid State Commun.* **148**, 538–540 (2008).

[7] Ogino, H. *et al.*, Superconductivity at 17 K in $(Fe_2P_2)(Sr_4Sc_2O_6)$: a new superconducting layered pnictide oxide with a thick perovskite oxide layer. *Supercond. Sci. Technol.* **22**, 075008(1-4) (2009).

[8] Zhu, X. *et al.*, Transition of stoichiometric $Sr_2VO_3FeAs$ to a superconducting state at 37.2 K. *Phys. Rev. B* **79**, 220512(1-4) (2009).

[9] Lee, C. H. *et al.*, Effect of Structural Parameters on Superconductivity in Fluorine-Free $LnFeAsO_{1-y}$ (Ln = La, Nd). *J. Phys. Soc. Jpn.* **77**, 083704(1-4) (2008).

[10] Mizuguchi, Y. *et al.*, Anion height dependence of $T_c$ for the Fe-based superconductor. *Supercond. Sci. Technol.* **23**, 054013(1-5) (2010).

[11] Kuroki, K. *et al.*, Pnictogen height as a possible switch between high-$T_c$ nodeless and low-$T_c$ nodal pairings in the iron-based superconductors. *Phys. Rev. B* **79**, 224511(1-16) (2009).

[12] Mizuguchi, Y. *et al.*, $BiS_2$-based layered superconductor $Bi_4O_4S_3$. *Phys. Rev. B* **86**, 220510(1-5) (2012).

[13] Mizuguchi, Y. *et al.*, Superconductivity in Novel $BiS_2$-Based Layered Superconductor $LaO_{1-x}F_xBiS_2$. *J. Phys. Soc. Jpn.* **81**, 114725(1-5) (2012).

[14] Usui, H., Suzuki, K. & Kuroki, K., Minimal electronic models for superconducting $BiS_2$ layers. *Phys. Rev. B* **86**, 220501(1-5) (2012).

[15] Xing, J. *et al.*, Superconductivity Appears in the Vicinity of an Insulating-Like Behavior in $CeO_{1-x}F_xBiS_2$. *Phys. Rev. B* **86**, 214518(1-5) (2012).

[16] Jha, R. *et al.*, Synthesis and superconductivity of new $BiS_2$ based superconductor $PrO_{0.5}F_{0.5}BiS_2$. *J. Supercond. Nov. Magn.* **26**, 499-502 (2013).

[17] Demura, S. *et al.*, $BiS_2$-based superconductivity in F-substituted $NdOBiS_2$. *J. Phys.*

*Soc. Jpn.* **82**, 033708(1-3) (2013).

[18] Thakur, G. S. *et al.*, Synthesis and properties of $SmO_{0.5}F_{0.5}BiS_2$ and enhancement in $T_c$ in $La_{1-y}Sm_yO_{0.5}F_{0.5}BiS_2$. Inorg. Chem. 54, 1076-1081 (2015).

[19] Yazici, D. *et al.*, Superconductivity of F-substituted $LnOBiS_2$ (Ln=La, Ce, Pr, Nd, Yb) compounds. *Philos. Mag.* **93**, 673(1-8) (2012).

[20] Okada, T. *et al.*, Topotactic Synthesis of a new $BiS_2$-based superconductor $Bi_2(O,F)S_2$. *Appl. Phys. Express* **8**, 023102(1-4) (2015).

[21] Shao, J. *et al.*, Superconductivity in $BiO_{1-x}F_xBiS_2$ and possible parent phase of $Bi_4O_4S_3$ superconductor. *Supercond. Sci. Technol.* **28**, 015008(1-6) (2015).

[22] Lin, X. *et al.*, Superconductivity induced by La doping in $Sr_{1-x}La_xFBiS_2$. *Phys. Rev. B* **87**, 020504(1-4) (2013).

[23] Jha, R., Tiwari, B. & Awana, V. P. S., Appearance of bulk superconductivity under hydrostatic pressure in $Sr_{0.5}RE_{0.5}FBiS_2$ (RE = Ce, Nd, Pr, and Sm) compounds. *J. Appl. Phys.* **117**, 013901(1-7) (2015).

[24] Zhai, H. F. *et al.*, Possible Charge-density wave, superconductivity and f-electron valence instability in $EuBiS_2F$. *Phys. Rev. B* **90**, 064518(1-9) (2014).

[25] Phelan, W. A. *et al.*, Stacking Variants and Superconductivity in the Bi-O-S System. *J. Am. Chem. Soc.* **135**, 5372-5374 (2013).

[26] Zhai, H. F. *et al.*, Anomalous Eu Valence State and Superconductivity in Undoped $Eu_3Bi_2S_4F_4$. *J. Am. Chem. Soc.* **136**, 15386-15393 (2014).

[27] Maziopa, A. K. *et al.*, Superconductivity in a new layered bismuth oxyselenide: $LaO_{0.5}F_{0.5}BiSe_2$. *J. Phys.: Condens. Matter* **26**, 215702(1-5) (2014).

[28] Deguchi, K. *et al.*, Evolution of superconductivity in $LaO_{1-x}F_xBiS_2$ prepared by high pressure technique. *EPL* **101**, 17004(p1-p5) (2013).

[29] Mizuguchi, Y. *et al.*, Stabilization of High-$T_c$ Phase of $BiS_2$-Based Superconductor $LaO_{0.5}F_{0.5}BiS_2$ Using High-Pressure Synthesis. *J. Phys. Soc. Jpn.* **83**, 053704(1-4) (2014).

[30] Kotegawa, H. *et al.*, Pressure Study of $BiS_2$-Based Superconductors $Bi_4O_4S_3$ and $La(O,F)BiS_2$. J. *Phys. Soc. Jpn.* **81**, 103702(1-4) (2012).

[31] Tomita, T. *et al.*, Pressure-Induced Enhancement of Superconductivity and Structural Transition in $BiS_2$-Layered $LaO_{1-x}F_xBiS_2$. *J. Phys. Soc. Jpn.* **83**, 063704(1-4) (2014).

[32] Wolowiec, C. T. *et al.*, Enhancement of superconductivity near the pressure-induced semiconductor–metal transition in the $BiS_2$-based superconductors $LnO_{0.5}F_{0.5}BiS_2$ (Ln = La, Ce, Pr, Nd). *J. Phys.: Condens. Matter* **25**, 42220(1-6) (2013).

[33] Kajitani, J. *et al.*, Chemical pressure effect on superconductivity of $BiS_2$-based

$Ce_{1-x}Nd_xO_{1-y}F_yBiS_2$ and $Nd_{1-z}Sm_zO_{1-y}F_yBiS_2$. *J. Phys. Soc. Jpn.* **84**, 044712(1-6) (2015).

[34] Hiroi, T. *et al.*, Evolution of superconductivity in $BiS_2$-based superconductor $LaO_{0.5}F_{0.5}Bi(S_{1-x}Se_x)_2$. *J. Phys. Soc. Jpn.* **84**, 024723(1-4) (2015).

[35] Miura, A. *et al.*, Structure, Superconductivity, and Magnetism of $Ce(O,F)BiS_2$ Single *Crystals. Cryst. Growth Des.* **15**, 39–44 (2015).

[36] Tanaka, M. *et al.*, Site Selectivity on Chalcogen Atoms in Superconducting La(O,F)BiSSe. *Appl. Phys. Lett.* **106**, 112601(1-5) (2015).

[37] Miura, A. *et al.*, Crystal structures of $LaO_{1-x}F_xBiS_2$ (x ~ 0.23, 0.46): effect of F doping on distortion of Bi-S plane. *J. Solid State Chem.* **212**, 213-217 (2014).

[38] Nagao, M. *et al.*, Structural Analysis and Superconducting Properties of F-Substituted $NdOBiS_2$ Single Crystals. *J. Phys. Soc. Jpn.* **82**, 113701(1-4) (2013).

[39] Tanaka, M. *et al.*, First single crystal growth and structural analysis of superconducting layered bismuth oxyselenide; $La(O,F)BiSe_2$. *J. Solid State Chem.* **219**, 168-172 (2014).

[40] Bardeen, J., Cooper, L. N. & Schrieffer, J. R., Theory of Superconductivity. *Phys. Rev.* **108**, 1175-1204 (1957).

[41] Mizuguchi, Y. *et al.*, Anisotropic upper critical field of $BiS_2$-based superconductor $LaO_{0.5}F_{0.5}BiS_2$. *Phys. Rev. B* **89**, 174515(1-7) (2014).

[42] Izumi, F. & Momma, Three-Dimensional Visualization in Powder Diffraction. *Solid State Phenom.* **130**, 15-20 (2007).

[43] Momma, K. & Izumi, F., VESTA: a three-dimensional visualization system for electronic and structural analysis. *J. Appl. Crystallogr.* **41**, 653-658 (2008).

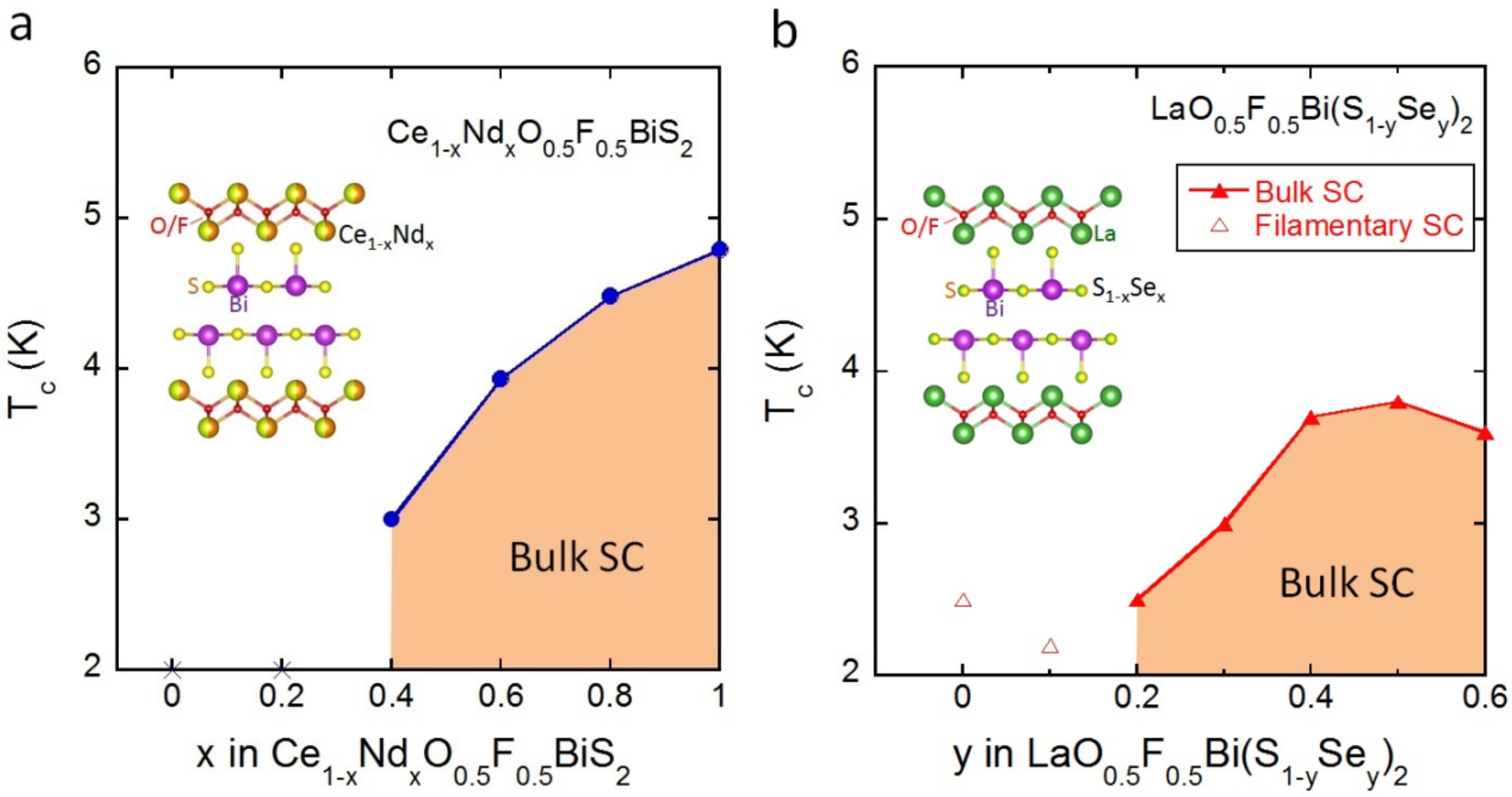

**Fig. 1. Superconductivity phase diagrams of $Ce_{1-x}Nd_xO_{0.5}F_{0.5}BiS_2$ and $LaO_{0.5}F_{0.5}Bi(S_{1-y}Se_y)_2$.**

**a.** Superconductivity phase diagrams of $Ce_{1-x}Nd_xO_{0.5}F_{0.5}BiS_2$. For $x = 0$ and 0.2, superconducting transition is not observed at $T > 2$ K. For $0.4 \le x \le 1$, superconducting transition with a large shielding signal, with which we could regard the samples as a bulk superconductor (Bulk SC), is observed. $T_c$ increases with increasing $x$. An inset figure shows a schematic image of crystal structure of $Ce_{1-x}Nd_xO_{0.5}F_{0.5}BiS_2$.

**b.** Superconductivity phase diagrams of $LaO_{0.5}F_{0.5}Bi(S_{1-y}Se_y)_2$. For $y = 0$ and 0.1, superconducting transition is observed but their shielding signals are very small as a bulk superconductor (Filamentary SC). For $y \ge 0.2$, superconducting transition with a large shielding signal is observed (Bulk SC). $T_c$ increases with increasing $y$ up to $y = 0.5$. An inset figure shows a schematic image of crystal structure of $LaO_{0.5}F_{0.5}Bi(S_{1-y}Se_y)_2$.

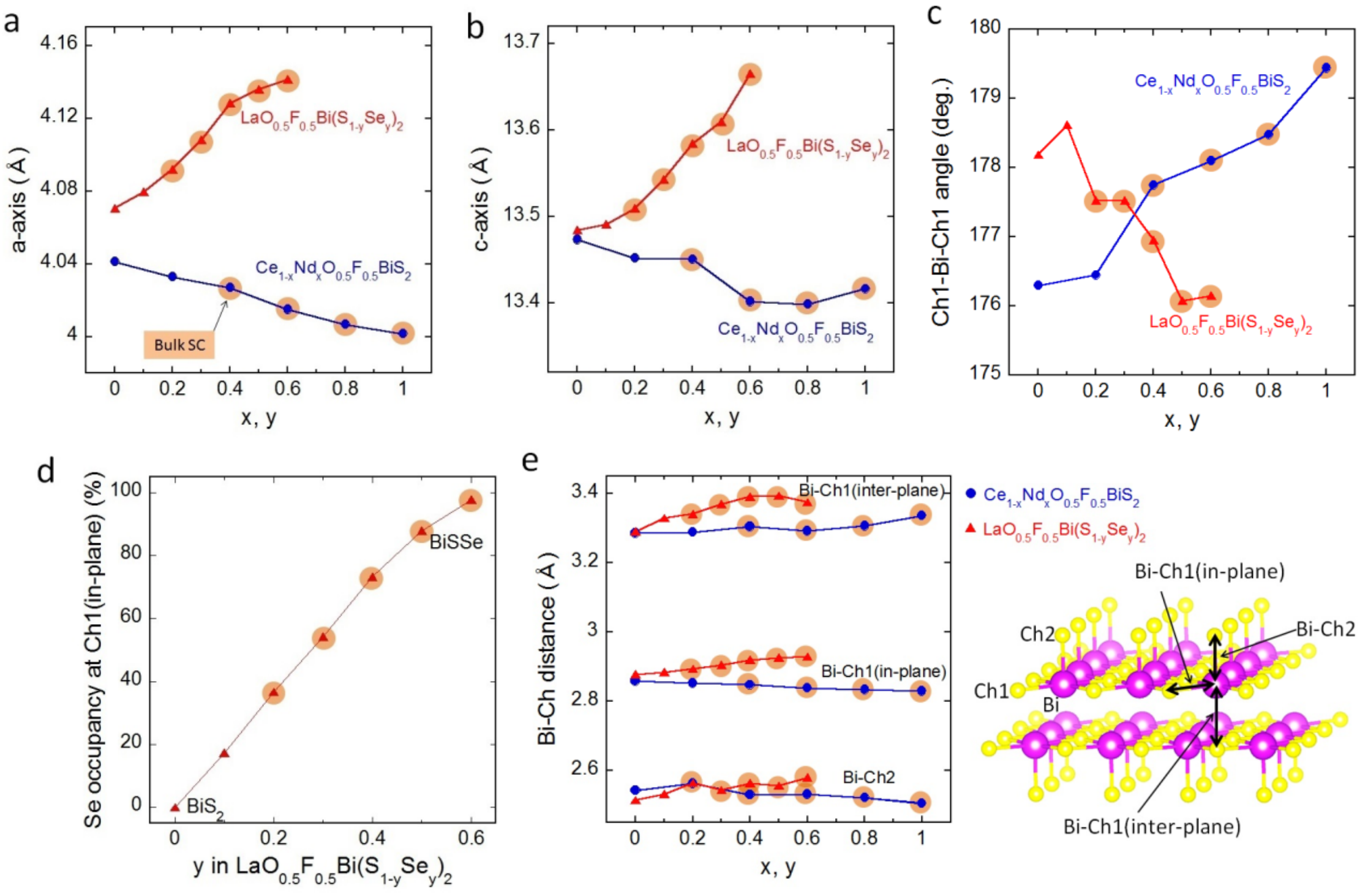

**Fig. 2. Crystal structure parameters for $Ce_{1-x}Nd_xO_{0.5}F_{0.5}BiS_2$ and $LaO_{0.5}F_{0.5}Bi(S_{1-y}Se_y)_2$.**

**a.** Lattice constant of *a* for $Ce_{1-x}Nd_xO_{0.5}F_{0.5}BiS_2$ and $LaO_{0.5}F_{0.5}Bi(S_{1-y}Se_y)_2$ is plotted as a function of *x* (or *y*). Data points for the samples showing bulk superconductivity (Bulk SC in Fig. 1) are highlighted with orange-filled circles in Figs. 2a-2e.

**b.** Lattice constant of *c* for $Ce_{1-x}Nd_xO_{0.5}F_{0.5}BiS_2$ and $LaO_{0.5}F_{0.5}Bi(S_{1-y}Se_y)_2$ is plotted as a function of *x* (or *y*).

**c.** Evolution of Ch1-Bi-Ch1 angle for $Ce_{1-x}Nd_xO_{0.5}F_{0.5}BiS_2$ and $LaO_{0.5}F_{0.5}Bi(S_{1-y}Se_y)_2$ as a function of *x* (or *y*).

**d.** Se concentration (*y*) dependence of occupancy of Se at the Ch1 site in the Bi-Ch1 plane for $LaO_{0.5}F_{0.5}Bi(S_{1-y}Se_y)_2$.

**e.** Evolution of three kinds of Bi-Ch distance for $Ce_{1-x}Nd_xO_{0.5}F_{0.5}BiS_2$ and $LaO_{0.5}F_{0.5}Bi(S_{1-y}Se_y)_2$ as a function of *x* (or *y*). The right crystal structure image describes the Bi, Ch1 and Ch2 sites, and three Bi-Ch distances: Bi-Ch1 (in-plane), Bi-Ch1 (inter-plane), and Bi-Ch2.

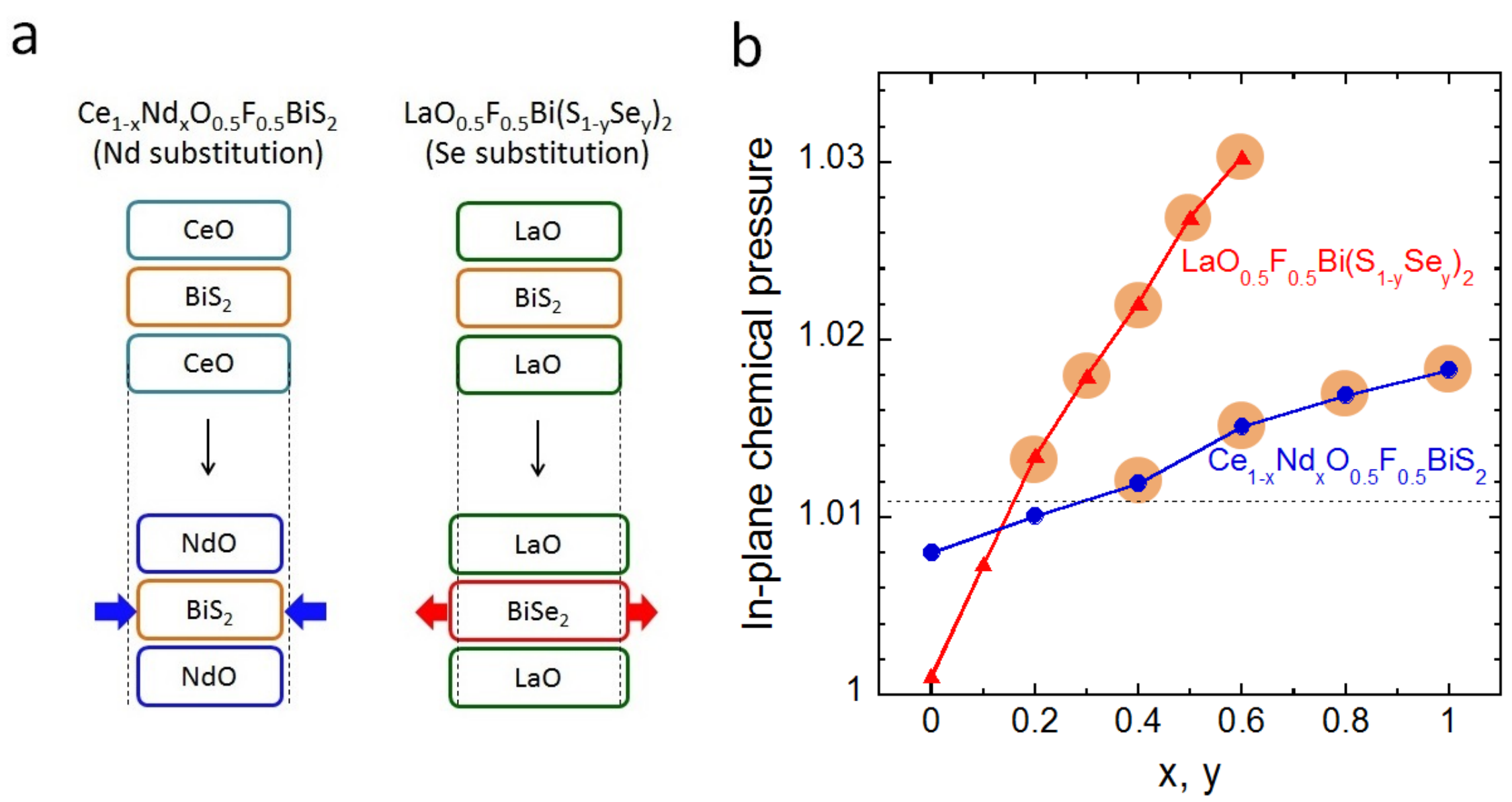

**Fig. 3. Effect of in-plane chemical pressure to superconductivity in $Ce_{1-x}Nd_xO_{0.5}F_{0.5}BiS_2$ and $LaO_{0.5}F_{0.5}Bi(S_{1-y}Se_y)_2$.**

**a.** Schematic image of changes in crystal structure with increasing in-plane chemical pressure in $Ce_{1-x}Nd_xO_{0.5}F_{0.5}BiS_2$ and $LaO_{0.5}F_{0.5}Bi(S_{1-y}Se_y)_2$. In $Ce_{1-x}Nd_xO_{0.5}F_{0.5}BiS_2$, a volume of spacer layer decreases with increasing Nd concentration ($x$), and Bi-S plane is compressed; hence, in-plane chemical pressure is enhanced. In $LaO_{0.5}F_{0.5}Bi(S_{1-y}Se_y)_2$, a volume of superconducting Bi-Ch1 layer increases with increasing Se concentration ($y$). However, the expansion of Bi-Ch1 plane is smaller than that expected from the the ionic radius of $S^{2-}$ and $Se^{2-}$ because the component of the spacer layer (LaO layer) does not change; hence, in-plane chemical pressure is enhanced as well as in $Ce_{1-x}Nd_xO_{0.5}F_{0.5}BiS_2$.

**b.** In-plane chemical pressure for $Ce_{1-x}Nd_xO_{0.5}F_{0.5}BiS_2$ and $LaO_{0.5}F_{0.5}Bi(S_{1-y}Se_y)_2$, calculated using the equation (1), are plotted as a function of $x$ (or $y$). In both systems, bulk superconductivity is induced with increasing chemical pressure. The dashed line at an in-plane chemical pressure of ~1.011 is an estimated boundary of bulk-SC and non-SC regions.

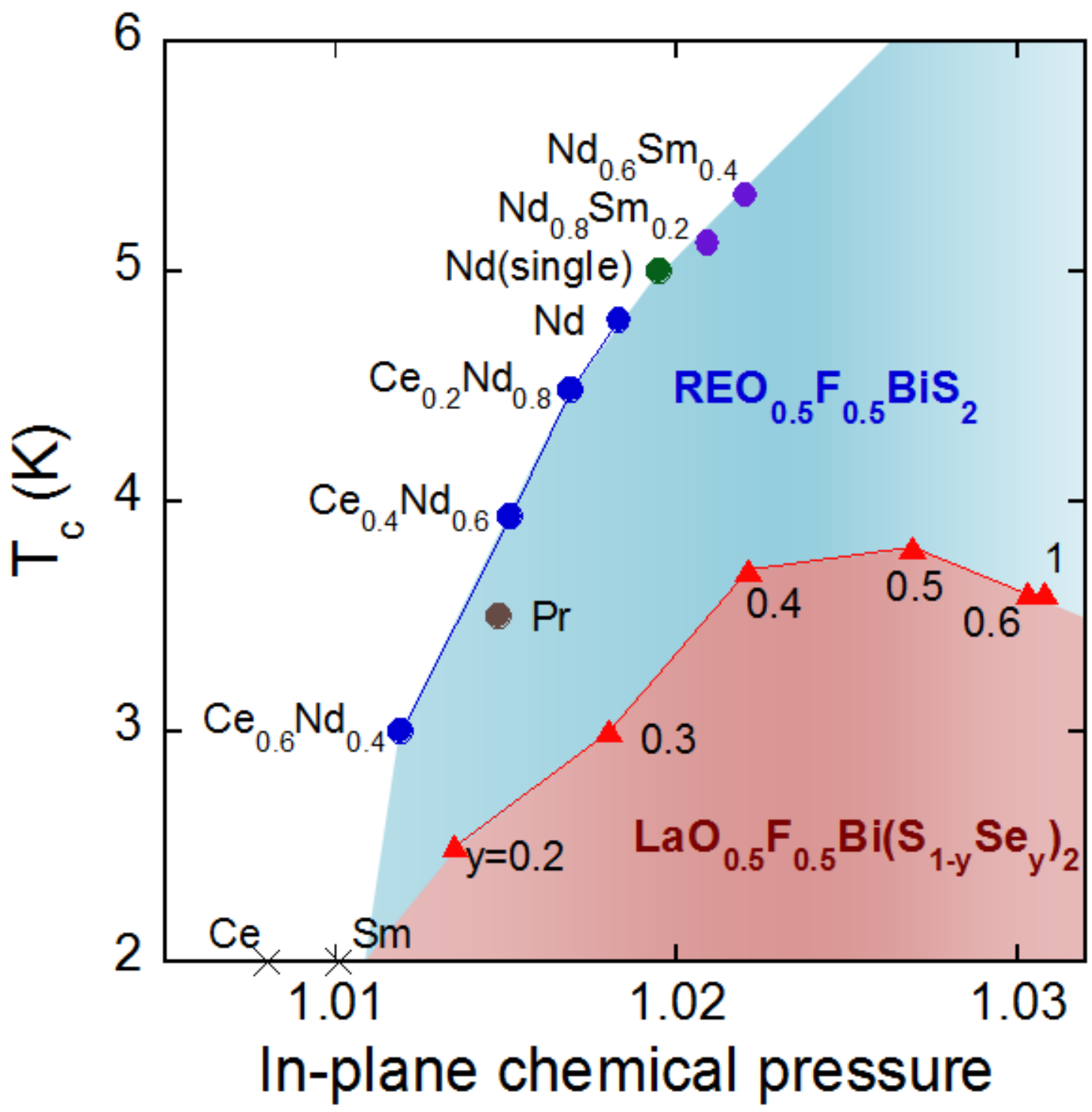

**Fig. 4. Relationship between $T_c$ and degree of in-plane chemical pressure in $REO_{0.5}F_{0.5}BiCh_2$.**

The data of $T_c$ and in-plane chemical pressure for $Ce_{1-x}Nd_xO_{0.5}F_{0.5}BiS_2$ and $LaO_{0.5}F_{0.5}Bi(S_{1-y}Se_y)_2$ are plotted with those for $Nd_{0.8}Sm_{0.2}O_{0.5}F_{0.5}BiS_2$ and $Nd_{0.6}Sm_{0.4}O_{0.5}F_{0.5}BiS_2$ [analyzed in this work], $NdO_{0.5}F_{0.5}BiS_2$ single crystal [38], $PrO_{0.5}F_{0.5}BiS_2$ [16], $SmO_{0.5}F_{0.5}BiS_2$ single crystal [18], and $LaO_{0.5}F_{0.5}BiSe_2$ ($y$ = 1) single crystal [39]. The data points of $REO_{0.5}F_{0.5}BiS_2$ and those of $LaO_{0.5}F_{0.5}Bi(S_{1-y}Se_y)_2$ exhibit different curves as separated by blue and red regions, respectively.

## Supplementary information

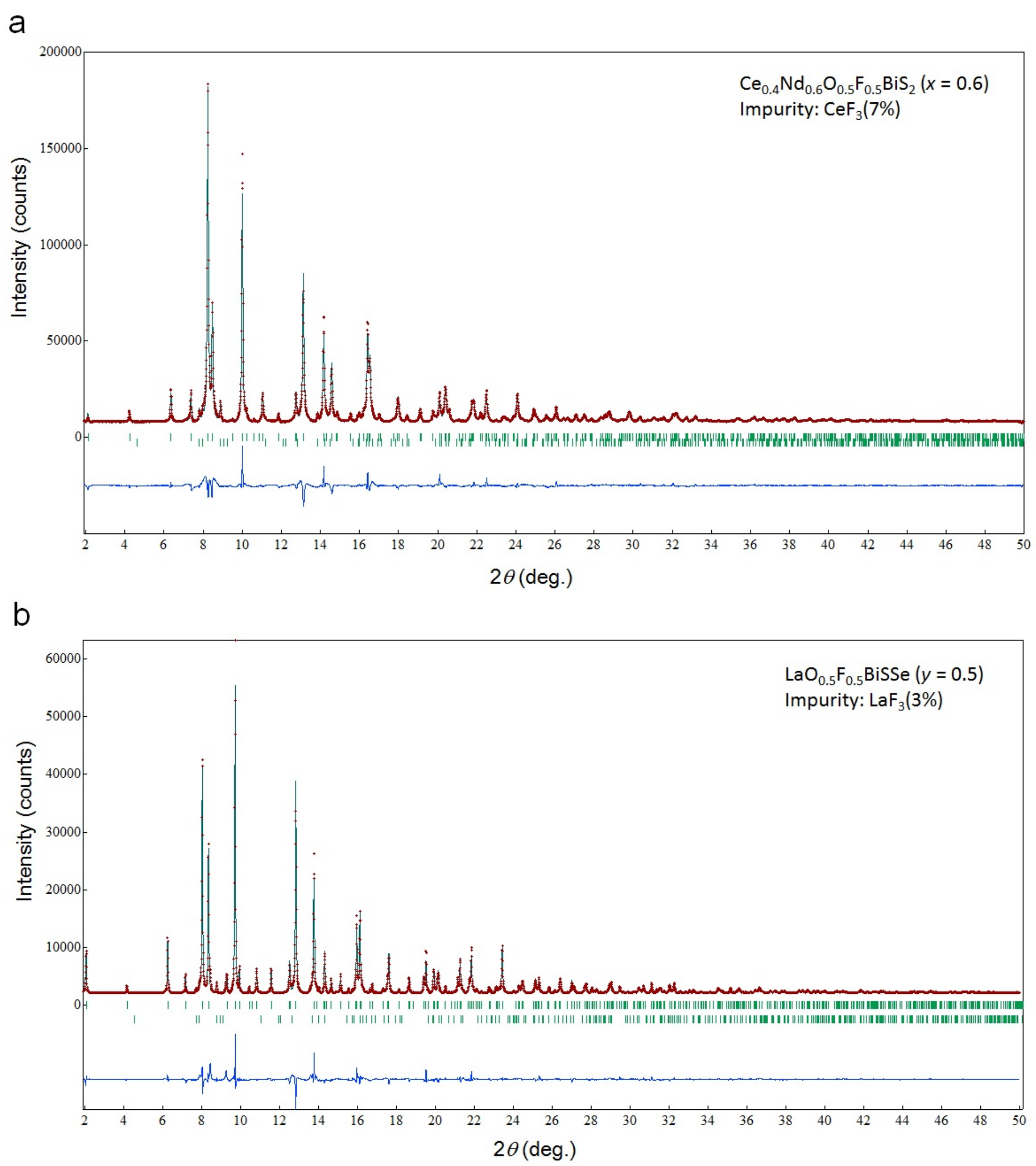

**Fig. S1. Typical XRD patterns for $Ce_{1\text{-}x}Nd_xO_{0.5}F_{0.5}BiS_2$ and $LaO_{0.5}F_{0.5}Bi(S_{1\text{-}y}Se_y)_2$.**

**a.** XRD patterns for $Ce_{0.4}Nd_{0.6}O_{0.5}F_{0.5}BiS_2$ ($x$ = 0.6). The Rietveld refinement was performed including a $CeF_3$ impurity phase (7 %). The upper and lower bars indicate calculated peak positions for $Ce_{0.4}Nd_{0.6}O_{0.5}F_{0.5}BiS_2$ and $CeF_3$, respectively. The reliable factor $R_{wp}$ was 4.9%.

**b.** XRD patterns for $LaO_{0.5}F_{0.5}BiSSe$ ($y$ = 0.5). The Rietveld refinement was performed including a $LaF_3$ impurity phase (3 %). The upper and lower bars indicate calculated peak positions for $LaO_{0.5}F_{0.5}BiSSe$ and $LaF_3$, respectively. The reliable factor $R_{wp}$ was 5.1%.

**Table S1. Obtained crystal structure parameters.**

Lattice constants *a* and *c*, *z* coordinate, occupancy of Se at Ch1 site ($g$(Se1)), and reliable factor of Rietveld refinement ($R_{wp}$) for $Ce_{1-x}Nd_xO_{0.5}F_{0.5}BiS_2$, $LaO_{0.5}F_{0.5}Bi(S_{1-y}Se_y)_2$, $Nd_{0.8}Sm_{0.2}O_{0.5}F_{0.5}BiS_2$, and $Nd_{0.6}Sm_{0.4}O_{0.5}F_{0.5}BiS_2$ are listed. Atomic coordinate for the $REO_{0.5}F_{0.5}BiS_2$ series are RE(0, 0.5, *z*), Bi(0, 0.5, *z*), Ch1(0, 0.5, *z*) , Ch2(0, 0.5, *z*), and O/F(0, 0, 0).

| Composition | *a* (Å) | *c* (Å) | *z*(RE) | *z*(Bi) | *z*(Ch1) | *z*(Ch2) | g(Se1) (%) | $R_{wp}$ (%) | Impurities |
|---|---|---|---|---|---|---|---|---|---|
| $CeO_{0.5}F_{0.5}BiS_2$ | 4.04119(7) | 13.4736(3) | 0.0975(1) | 0.6249(1) | 0.3810(6) | 0.8136(5) | - | 5.8 | $CeF_3$(9%) |
| $Ce_{0.8}Nd_{0.2}O_{0.5}F_{0.5}BiS_2$ | 4.0330(1) | 13.4522(6) | 0.0976(2) | 0.6255(2) | 0.3810(8) | 0.8160(8) | - | 5.8 | $CeF_3$(9%) |
| $Ce_{0.6}Nd_{0.4}O_{0.5}F_{0.5}BiS_2$ | 4.02710(8) | 13.4510(8) | 0.0965(1) | 0.6249(1) | 0.3792(6) | 0.8131(5) | - | 4.9 | $CeF_3$(8%) |
| $Ce_{0.4}Nd_{0.6}O_{0.5}F_{0.5}BiS_2$ | 4.01501(9) | 13.4023(4) | 0.0977(1) | 0.6246(1) | 0.3789(6) | 0.8135(6) | - | 4.9 | $CeF_3$(7%) |
| $Ce_{0.2}Nd_{0.8}O_{0.5}F_{0.5}BiS_2$ | 4.00682(9) | 13.3997(4) | 0.0969(1) | 0.6248(1) | 0.3780(7) | 0.8130(6) | - | 5.9 | $CeF_3$(3%) + $NdF_3$(2%) |
| $NdO_{0.5}F_{0.5}BiS_2$ | 4.0017(1) | 13.4166(5) | 0.0973(2) | 0.6249(2) | 0.3762(9) | 0.8117(8) | - | 7.4 | $Bi_2S_3$(2%) + $NdF_3$(5%) |
| $Nd_{0.8}Sm_{0.2}O_{0.5}F_{0.5}BiS_2$ | 3.99205(9) | 13.4152(4) | 0.0956(1) | 0.6246(1) | 0.3763(8) | 0.8084(6) | - | 7.6 | unknown |
| $Nd_{0.6}Sm_{0.4}O_{0.5}F_{0.5}BiS_2$ | 3.9880(2) | 13.4407(6) | 0.0963(2) | 0.6253(2) | 0.377(1) | 0.8088(8) | - | 9.2 | unknown |
| $LaO_{0.5}F_{0.5}BiS_2$ | 4.07063(6) | 13.4848(3) | 0.0978(1) | 0.6237(1) | 0.3797(6) | 0.8103(5) | 0.0 | 5.8 | $Bi_2S_3$(1%) + $LaF_3$(7%) |
| $LaO_{0.5}F_{0.5}BiS_{1.8}Se_{0.2}$ | 4.07966(3) | 13.4914(2) | 0.09756(9) | 0.62469(7) | 0.3779(4) | 0.8124(4) | 17.3(5) | 5.4 | $LaF_3$(4%) |
| $LaO_{0.5}F_{0.5}BiS_{1.6}Se_{0.4}$ | 4.09211(8) | 13.5096(3) | 0.0971(2) | 0.6260(2) | 0.3786(5) | 0.8159(7) | 36.7(8) | 5.6 | $LaF_3$(9%) |
| $LaO_{0.5}F_{0.5}BiS_{1.4}Se_{0.6}$ | 4.10821(4) | 13.5433(2) | 0.0964(1) | 0.62673(9) | 0.3779(3) | 0.8146(4) | 54.2(5) | 5.4 | $LaF_3$(4%) |
| $LaO_{0.5}F_{0.5}BiS_{1.2}Se_{0.8}$ | 4.12832(4) | 13.5848(2) | 0.0959(1) | 0.62769(8) | 0.3780(2) | 0.8165(4) | 73.4(5) | 5.5 | $LaF_3$(3%) |
| $LaO_{0.5}F_{0.5}BiSSe$ | 4.13593(5) | 13.6098(2) | 0.0953(1) | 0.6284(1) | 0.3790(2) | 0.8163(4) | 87.9(5) | 5.1 | $LaF_3$(3%) |
| $LaO_{0.5}F_{0.5}BiS_{0.8}Se_{1.2}$ | 4.14138(7) | 13.6656(3) | 0.0944(1) | 0.6271(1) | 0.3801(3) | 0.8160(4) | 97.9(7) | 6.0 | unknown |